%Paper: hep-th/9207060
%From: <JIAN%TAMPHYS.BITNET@ricevm1.rice.edu>
%Date: Thu, 16 Jul 92 11:29 CST

%plain tex
\magnification 1200
\baselineskip=18pt
\overfullrule=0pt

\overfullrule=0pt
\def\dg{\hbox{$^\dagger$}}

\def\ddg{\hbox{$^\ddagger$}}
\def\lam{\hbox{$\lambda\kern-6pt^{\_\_}$}}

\def\r{\vbox{\hbox{\raise1.5mm\hbox{$>$}}
\kern-18pt\hbox{\lower1.5mm\hbox{$\sim$}}}}
\def\l{\vbox{\hbox{\raise1.5mm\hbox{$<$}}
\kern-18pt\hbox{\lower1.5mm\hbox{$\sim$}}}}

\def\leaderfill{\leaders\hbox to 1em{\hss.\hss}\hfill}
\def\rs{\vbox{\hbox{\raise1.1mm\hbox{$>$}}
\kern-18pt\hbox{\lower1.1mm\hbox{$\sim$}}}}
\def\ls{\vbox{\hbox{\raise1.1mm\hbox{$<$}}
\kern-18pt\hbox{\lower1.1mm\hbox{$\sim$}}}}

\line{\hfil CERN-TH.6560/92}
\line{\hfil CTP/TAMU-37/92}
\vskip .50truein
\centerline{\bf TYPE II P-BRANES:  THE BRANE-SCAN REVISITED}
\vskip .75truein
\centerline{M. J. Duff{\footnote\dg{Work supported in part by NSF grant
PHY-9106593}} and J. X. Lu{\footnote\ddg{Supported by a World Laboratory
Scholarship}}}
\vskip .75truein
\centerline{Theory Division, CERN, Geneva}
\centerline{and}
\centerline{Center for Theoretical Physics}
\centerline{Physics Department}
\centerline{Texas A\&M University}
\centerline{College Station, TX  77843}
\vskip .75truein
\centerline{\bf ABSTRACT}
\medskip
We re-examine the classification of supersymmetric extended objects in the
light of the recently discovered Type II $p$-branes, previously thought not to
exist for $p > 1$.  We find new points on the brane-scan only in $D = 10$ and
then only for $p = 3$ (Type IIB), $p = 4$ (Type IIA), $p = 5$ (Types IIA and
IIB) and $p = 6$ (Type IIA).  The case $D = 10$, $p = 2$ (Type IIA) also
exists but is equivalent to the previously classified $D = 11$ supermembrane.
\bigskip

\noindent
CERN-TH.6530/92

\noindent
July 1992
\vfill\eject

\noindent
{\bf 1.  Introduction}
\medskip
In 1986, Hughes et al [1] discovered a superthreebrane with worldvolume
dimension $d = 4$ in $D = 6$ spacetime.  Their superthreebrane action was a
generalization of the Green-Schwarz action for superstrings in that it
exhibited spacetime supersymmetry and worldvolume fermionic $\kappa$-symmetry.
Shortly afterwards, Bergshoeff, Sezgin and Townsend [2] found corresponding
actions for other values of $d$~and~$D$, called super ``$p$-branes'' where $p =
d - 1$ is the number of spatial dimensions of the worldvolume.  Moreover, Duff,
Howe, Inami and Stelle [3] showed how the action for a super ($p - 1$)-brane in
($D - 1$) dimensions could be derived from that of a $p$-brane in $D$
dimensions by the process of simultaneous dimensional reduction.  A complete
classification of all supersymmetric extended objects, incorporating all of the
previous observations, was then attempted by Achucarro, Evans, Townsend and
Wiltshire [4].  Their results, which we shall discuss in section 2, can be
summarized by the ``brane-scan'' of Figure 1.  According to this
classification, Type II $p$-branes, i.e those with $N = 2$ spacetime
supersymmetry, do not exist for $p > 1$.

Recently, however, it has been discovered that, in $D = 10$, super $p$-branes
exist not only for Type IIA and IIB strings ($p = 1$) but also for Type IIA and
IIB fivebranes ($p = 5$) [5] and Type IIB threebranes ($p = 3$) [6].  The no-go
theorem is circumvented because in addition to the superspace coordinates
$X^M$~and~$\theta^{\alpha}$ there are also higher spin fields on the world
volume:  vectors or antisymmetric tensors.  This raises the question:  are
there other Type II super $p$-branes, and if so, for what $p$~and~$D$?  The
purpose of the present paper will be to attempt to classify these new
supersymmetric extended objects.

We begin in section 3 by asking what new points on the brane-scan are
permitted by bose-fermi matching alone.  There are surprisingly few:  $p = 3,
4, \ldots 9$ in $D = 10$; $p = 3, 4, 5$ in $D = 6$ and $p = 3$ in $D = 4$.  The
much harder task is to narrow down these possibilities to objects that actually
exist.  One obvious handicap is that, unlike the p-branes discussed by
Achucarro et al [4], no-one has yet succeeded in writing down the action for
these new Type II $p$-branes.  The existence of the $p = 3$~and~$p = 5$ objects
mentioned above was established indirectly:  by showing that they emerge as
soliton solutions of either Type IIA or Type IIB supergravity.  The nature of
the worldvolume fields is then established by studying the zero modes of the
soliton.  In particular, a super $p$-brane requires that the soliton solution
preserves some unbroken supersymmetry and hence that the zero modes form a
supermultiplet.  Although we know of no general proof that all supersymmetric
extended objects correspond to a soliton, this is true of all those on the old
brane-scan and thus seems a good guide to constructing the new one.  Following
this route we shall conclude that of all the possible $D = 10$ Type II super
$p$-branes permitted by bose-fermi matching alone, only those with $p = 0$
(Type IIA), $p = 1$ (Type IIA and IIB), $p = 3$ (Type IIB), $p = 4$ (Type IIA)
$p = 5$ (Type IIA and IIB) and $p = 6$ (Type IIA) actually exist.  [The reader
may wonder why there seems to be a gap at $p = 2$.  Indeed, duality would seem
to demand that in $D = 10$ a Type IIA superfourbrane should imply a Type IIA
supermembrane.  This object does indeed exist but it should not be counted as a
new theory since vectors are dual to scalars in $d = 3$ and so its worldvolume
action is simply obtained by dualizing one of the 11 $X^M$ of the $D = 11$
supermembrane.]  Our results thus confirm the conjecture of Horowitz and
Strominger [8] that super Type II $p$-brane solitons in $D = 10$ exist for all
$0 \leq p \leq 6$.  Moreover, we find that no new fundamental Type II $p$-
branes emerge in $D < 10$.
\medskip

\noindent
{\bf 2.  The old brane-scan}
\medskip

As the $p$-brane moves through spacetime, its trajectory is described by the
functions $X^M (\xi)$ where $X^M$ are the spacetime coordinates ($M = 0, 1,
\ldots, D - 1$) and $\xi^i$ are the worldvolume coordinates ($i = 0, 1, \ldots,
d - 1$).  It is often convenient to make the so-called ``static gauge choice''
by making the $D = d + (D - d)$ split

$$X^M (\xi) = (X^{\mu} (\xi), Y^m (\xi))\eqno(2.1)$$

\noindent
where $\mu = 0, 1, \ldots, d - 1$~and~$m = d, \ldots, D - 1$, and then setting

$$X^{\mu} (\xi) = \xi^{\mu}\eqno(2.2)$$

\noindent
Thus the only physical worldvolume degrees of freedom are given by the $(D -
d)~Y^m (\xi)$.  So the number of on-shell bosonic degrees of freedom is

$$N_B = D - d\eqno(2.3)$$

\noindent
To describe the super $p$-brane we augment the $D$ bosonic coordinates $X^M
(\xi)$ with anticommuting fermionic coordinates $\theta^{\alpha} (\xi)$.
Depending on $D$, this spinor could be Dirac, Weyl, Majorana or Majorana-Weyl.
The fermionic $\kappa$-symmetry means that half of the spinor degrees of
freedom are redundant and may be eliminated by a physical gauge choice.  The
net result is that the theory exhibits a {\it $d$-dimensional worldvolume
supersymmetry} where the number of fermionic generators is exactly half of the
generators in the original spacetime supersymmetry.  This partial breaking of
supersymmetry is a key idea.  Let $M$ be the number of real components of the
minimal spinor and $N$ the number of supersymmetries in $D$ spacetime
dimensions and let $m$~and~$n$ be the corresponding quantities in $d$
worldvolume dimensions.  Let us first consider $d > 2$.  Since $\kappa$-
symmetry always halves the number of fermionic degrees of freedom and going
on-shell halves it again, the number of on-shell fermionic degrees of freedom
is

$$N_F = {1\over 2}~mn = {1\over 4}~MN\eqno(2.4)$$

\noindent
Worldvolume supersymmetry demands $N_B = N_F$ and hence

$$D - d = {1\over 2}~mn = {1\over 4}~MN\eqno(2.5)$$

\noindent
A list of dimensions, number of real components of the minimal spinor and
possible supersymmetries is given in Table 1, from which we see that there are
only 8 solutions to (2.5) all with $N = 1$, as shown in Fig. 1.  We note in
particular that $D_{{\rm max}} = 11$ since $M \geq 64$ for $D \geq 12$ and
hence (2.5) cannot be satisfied.  Similarly $d_{{\rm max}} = 6$ since $m \geq
16$ for $d \geq 7$.  The case $d = 2$ is special because of the ability to
treat left and right moving modes independently.  If we require the sum of both
left and right moving bosons and fermions to be equal, then we again find the
condition (2.5).  This provides a further 4 solutions all with $N = 2$,
corresponding to Type II superstrings in $D = 3, 4, 6$~and~$10$ (or 8 solutions
in all if we treat Type IIA and Type IIB separately.  The gauge-fixed Type IIB
superstring will display (8, 8) supersymmetry on the worldsheet and the Type
IIA will display (16, 0), the opposite [5] of what one might naively expect).
If we require only left (or right) matching, then (2.5) is replaced by

$$D - 2 = n = {1\over 2}~MN\eqno(2.6)$$

\noindent
which allows another 4 solutions in $D = 3, 4, 6$~and~$10$, all with $N = 1$.
The gauge-fixed theory will display (8,0) worldsheet supersymmetry.  The
heterotic string falls into this category.  The results are also shown in Fig.
1.

An equivalent way to arrive at the above conclusions is to list all scalar
supermultiplets in $d \geq 2$ dimensions and to interpret the dimension of the
target space, $D$, by

$$D - d =~{\rm number~of~scalars}\eqno(2.7)$$

\noindent
A useful reference is Strathdee [9] who provides an exhaustive classification
of all unitary representations of supersymmetry with maximum spin 2.  In
particular, we can understand $d_{{\rm max}} = 6$ from this point of view since
this is the upper limit for scalar supermultiplets.  In summary, according to
the above classification, Type II $p$-branes do not exist for $p > 1$.
\bigskip

\halign{\indent #&\qquad\hfil# \hfil&\quad\hfil #\hfil&\quad\hfil # \hfil &
\quad \hfil # \hfil &\quad \hfil # \hfil &\quad #\hfil\cr
&&&Dimension & Minimal Spinor& Supersymmetry&\cr
&&&($D$ or $d$) & ($M$ or $m$) & ($N$ or $n$)&\cr
&&&11 & 32 & 1&\cr
&&&10 & 16 & 2, 1&\cr
&&&9 & 16 & 2, 1&\cr
&&&8 & 16 &2, 1&\cr
&&&7 & 16 & 2, 1&\cr
&&&6 & 8 & 4, 3, 2, 1&\cr
&&&5 & 8 & 4, 3, 2, 1&\cr
&&&4 & 4 & 8, $\ldots$, 1&\cr
&&&3 & 2 & 16, $\ldots$, 1&\cr
&&&2 & 1 & 32, $\ldots$, 1&\cr}

\noindent
Table 1.  Minimal spinor components and supersymmetries.
\vfill\eject

\noindent
{\bf 3.  Bose-fermi matching:  a necessary condition}
\medskip

Given that the gauge-fixed theories display worldvolume supersymmetry, and
given that we now wish to include the possibility of vector (and/or
antisymmetric tensor) fields, it is a relatively straightforward exercise to
repeat the bose-fermi matching conditions for vector (and/or antisymmetric
tensor) supermultiplets.  Once again, we may proceed in one of two ways.
First, given that a worldvolume vector has ($d - 2$) degrees of freedom, the
scalar multiplet condition (2.5) gets replaced by

$$D - 2 = {1\over 2}~mn = {1\over 4}~MN\eqno(3.1)$$

\noindent
Alternatively, we may simply list all the supermultiplets in Strathdee's
classification and once again interpret $D$ via (2.7).  The results are shown
in Fig. 2.

Several comments are now in order:

\item{1)} Vector supermultiplets exist only for $4 \leq d \leq 10$ [9].  In $d
= 3$ vectors have only 1 degree of freedom and are dual to scalars.  So these
multiplets will already have been included as scalar multiplets in section 2.
In $d = 2$, vectors have no degrees of freedom.

\item{2)} The number of scalars in a vector supermultiplet is such that, from
(2.7), $D = 4, 6$ or $10$ only, in accordance with (3.1).

\item{3)} Repeating the analysis for antisymmetric tensors does not introduce
any new points on the scan.  For example in $d = 6$ there is a chiral (2, 0)
tensor supermultiplet, with a second rank tensor whose field strength is self
dual:  $(B_{\mu\nu}^-, \lambda^I, \phi^{[IJ]})$, $I = 1, \ldots, 4$,
corresponding to the Type IIA fivebrane and a non-chiral (1, 1) vector
multiplet $(B_{\mu}, \chi^I, A^J\,_I, \xi)$, $I = 1, 2$, corresponding to the
Type IIB fivebrane [5].  Both occupy the $(d = 6, D = 10)$ slot in Fig. 2.

\item{4)} We emphasize that Fig. 2 merely tells us what is allowed by
bose/fermi matching.  We must now try to establish which of these possibilities
actually exists.
\medskip

\noindent
{\bf 4.  $p$-brane solutions}
\medskip

All of the circles on the brane-scan are known to correspond to soliton
solutions of an underlying supersymmetric field theory [1, 7, 10, 11, 12, 13,
14].  As for the crosses, supersymmetric soliton solutions of both Type IIA and
Type IIB supergravity have been found for the case $(d = 6, D = 10)$ [5] and of
Type IIB for $(d = 4, D = 10)$ [6].  What about the others?  In this section we
shall exhibit the solutions and then in section 5 ask whether they are
supersymmetric.

To this end, consider the following generic $D = 10$ action

$$I_{10} (d) = {1\over 2\kappa^2}~\int d^{10} x \sqrt{-g} \Bigg(R - {1\over
2}~(\partial \phi)^2 - {1\over 2 (d+1)!}~e^{-\alpha \phi}
F^2_{d+1}\Bigg)\eqno(4.1)$$

\noindent
This describes the interaction of an antisymmetric tensor potential of rank
$d$, $A_{M_1 M_2 \ldots M_d}\break\hfil (M = 0, 1, \ldots 9)$, interacting with
gravity $g_{MN}$, and the dilaton $\phi$, where the rank $(d+1)$ field strength
$F_{d+1}$ is given by

$$F_{d+1} = d A_d\eqno(4.2)$$

\noindent
and the constant $\alpha$ is given by

$$\alpha = {(4-d)\over 2}~(-1)^d~.\eqno(4.3)$$

\noindent
To solve the corresponding field equations, we follow [12, 13] and make an
ansatz corresponding to the most general $d/(10-d)$ split invariant under $P_d
\times SO (10-d)$ where $P_d$ is the $d$-dimensional Poincar\'e group.  (The
black $(d-1)$ branes discussed in section 5 exhibit $P_d$ invariance only in
the mass $=$ charge limit [8], so this ansatz will automatically single out
these extreme cases.)  We split the indices

$$x^M = (x^{\mu}, y^m)\eqno(4.4)$$

\noindent
where $\mu = 0, 1, \ldots (d - 1)$~and~$m = d, d + 1, \ldots 9$ and write the
line-element as

$$ds^2 = e^{2A} \eta_{\mu\nu} dx^{\mu} dx^{\nu} + e^{2B} \delta_{mn} dy^m
dy^n\eqno(4.5)$$

\noindent
and the $d$-form gauge field as

$$A_{01 \ldots d} = - e^C\eqno(4.6)$$

\noindent
All other components of $A_{M_1 \ldots M_d}$ are set to zero.  $P_d$ invariance
then requires that the arbitrary functions $A$, $B$, $C$ depend only on $y^m$;
$SO (10-d)$ invariance requires that this dependence be only through $r =
\sqrt{\delta_{mn} y^m y^n}$.  Similarly our ansatz for the dilaton is

$$\phi = \phi (r)\eqno(4.7)$$

Substituting these ansatze into the field equations leads to the following
solutions, assuming that $g_{MN}$ tends asymptotnally to $\eta_{MN}$:

$$\eqalign{A&={\tilde{d}\over 16}~C\cr
B&={-d\over 16}~C\cr
\phi&={\alpha\over 2}~C\cr}\eqno(4.8)$$

\noindent
where, for simplicity, we have set the vev of the dilaton equal to zero.  $C$
is given by

$$e^{-C} = 1 + {k_d\over r^{\tilde{d}}} \qquad \tilde{d} > 0\eqno(4.9)$$

\noindent
where $k_d$ is a constant.  Here we have introduced $\tilde{d}$, the dimension
of the extended object dual to the $(d - 1)$-brane in $D = 10$,

$$\tilde{d} = 8 - d\eqno(4.10)$$

\noindent
We shall refer to these solutions as ``elementary $(d - 1)$-branes''.  They are
characterized by a non-vanishing electric Noether charge

$$e_d = {1\over \sqrt{2}\kappa} \int\limits_{S^{\tilde{d} + 1}} e^{-
\alpha\phi}{}^{\ast}\!F = {1\over \sqrt{2}\kappa}~\tilde{d} \Omega_{\tilde{d} +
1} k_d\eqno(4.11)$$

\noindent
where $S^{\tilde{d} + 1}$ is the $(\tilde{d} + 1)$ sphere surrounding the
elementary $(d - 1)$ brane, and $\Omega_{\tilde{d} + 1}$ is volume of the unit
$S^{\tilde{d} + 1}$.  Strictly speaking, these configurations display $\delta
(r)$ singularities and fail to solve the field equations at $r = 0$ unless we
augment the action (4.1) by the action for the $(d - 1)$-brane source.  We need
not dwell on this here; a full discussion may be found in [15].

However, we can also find non-singular ``solitonic $(\tilde{d} - 1)$-brane''
solutions which are characterized by a non-vanishing topological magnetic
charge

$$g_{\tilde{d}} = {1\over \sqrt{2}\kappa}~\int\limits_{S^{d + 1}}
F\eqno(4.12)$$

\noindent
satisfying the Dirac quantization condition

$$e_d g_{\tilde{d}} = 2 \pi n \qquad n =~{\rm integer}\eqno(4.13)$$

\noindent
To obtain these solutions we now make an ansatz invariant under $P_{\tilde{d}}
\times SO (10 - \tilde{d})$.  Hence we write (4.4) and (4.5) as before where
now $\mu = 0, 1 \ldots (\tilde{d} - 1)$~and~$m = \tilde{d}, \tilde{d} + 1,
\ldots 9$.  The ansatz for the antisymmetric tensor, however, will be made on
the field strength rather than the potential.

$${1\over \sqrt{2}\kappa}~F_{d + 1} = g_{\tilde{d}} \varepsilon_{d +
1}/\Omega_{d + 1}\eqno(4.14)$$

\noindent
where $\varepsilon_{d + 1}$ is the volume form on $S^{d + 1}$.  Since this is a
harmonic form, $F$ can no longer be written globally as the curl of $A$, but it
satisfies the Bianchi identities.  It is now not difficult to show that all the
fields equations are satisfied simply by making the replacements $d \rightarrow
\tilde{d}$ and hence $\alpha (d) \rightarrow \alpha (\tilde{d}) = - \alpha (d)$
in (4.8-10).

One may now consider the theory ``dual'' to (4.1) for which the roles of
antisymmetric tensor field equations and Bianchi identities (and hence electric
and magnetic charges) are interchanged.  The action is given by

$$\tilde{I}_{10} (\tilde{d}) = {1\over 2\kappa^2}~\int d^{10} x \sqrt{-g}
\Bigg(R - {1\over 2}~(\partial \phi)^2 - {1\over 2 (\tilde{d} +
1)!}~e^{\alpha\phi} \tilde{F}^2_{\tilde{d} + 1}\Bigg)\eqno(4.15)$$

\noindent
where the rank $(\tilde{d} + 1)$ field strength $\tilde{F}$ is the curl of a
$\tilde{d}$-form potential

$$\tilde{F}_{\tilde{d}+1} = d \tilde{A}_{\tilde{d}}\eqno(4.16)$$

\noindent
and is related to $F$ via

$$\tilde{F}_{\tilde{d}+1} = e^{-\alpha\phi}{}^{\ast}\!F_{d+1}\eqno(4.17)$$

\noindent
$\alpha$ is the same constant appearing in (4.1) but occurs with the opposite
sign.  It should be clear that the system $\tilde{I}_{10} (\tilde{d})$ admits
the same elementary and solitonic solutions as $I_{10} (d)$ provided we
everywhere make the replacement $d \rightarrow \tilde{d}$ and hence $\alpha (d)
\rightarrow \alpha (\tilde{d}) = - \alpha (d)$.  In the dual theory, therefore,
the roles of elementary and solitonic solutions are interchanged.

Now let us return to the question of supersymmetry.  First of all, the generic
action (4.1) correctly describes the bosonic sector of the 3-form field
strength version of $N = 1$, $D = 10$ supergravity, the field theory limit of
the superstring.  We simply set $d = 2$ and hence $\tilde{d} = 6$~and~$\alpha
= 1$.  The resulting elementary solution (4.8-9) is the Dabholkar et al string
[10] and the soliton solution is the Duff-Lu fivebrane [13].  The action (4.1)
also describes the bosonic sector of the 7-form field strength version of $N =
1$, $D = 10$ supergravity.  We simply set $d = 6$ and hence $\tilde{d} =
2$~and~$\alpha = - 1$.  The resulting elementary solution (4.8-9) is the
Duff-Lu fivebrane [13], and the soliton solution is the Dabholkar et al [10]
string.  As shown in [10,13], both the string and the fivebrane break one half
of the spacetime supersymmetries.
\bigskip

\halign{\indent #&\qquad\hfil#\hfil&\quad\hfil # \hfil &
\quad \hfil # \hfil &\quad \hfil # \hfil &\quad \hfil# \hfil &
\quad \hfil # \hfil &\quad \hfil # \hfil&\quad #\hfil \cr
&&$d$ & $\tilde{d}$ & $\alpha (d)$ & A & B & $\phi$&\cr
&&1 & 7 & $-3/2$ & $7C/16$ & $-C/16$ & $-3C/4$&\cr
&&2 & 6 & 1 & $3C/8$ & $-C/8$ & $C/2$&\cr
&&3 & 5 & $-1/2$ & $5C/16$ & $-3C/16$ & $-C/4$&\cr
&&4 & 4 & 0 & $C/4$ & $-C/4$ & 0&\cr
&&5 & 3 & $1/2$ & $3C/16$ & $-5C/16$ & $C/4$&\cr
&&6 & 2 & $-1$ & $C/8$ & $-3C/8$ & $-C/2$&\cr
&&7 & 1 & $3/2$ & $C/16$ & $-7/16$ & $3C/4$&\cr}

\noindent
Table 2.  The functions A, B and $\phi$ in terms of C as demanded
by supersymmetry.
\medskip

Now let us turn to $D = 10$ Type IIA supergravity, whose bosonic action is
given by

$$\eqalign{I_{10} (IIA)&={1\over 2\kappa^2}~\int d^{10} x \sqrt{-g}
\Bigg[R - {1\over 2}~(\partial\phi)^2 - {1\over 2.3!}~e^{-\phi} F_3\,^2\cr
&-{1\over 2.2!}~e^{3 \phi/2} F_2\,^2 - {1\over 2.4!}~e^{\phi/2}
F'_4\,^2\Bigg]\cr
&- {1\over 8\kappa^2} \int F_4 \wedge F_4 \wedge A_2\cr}\eqno(4.18)$$

\noindent
where

$$F'_4 = dA_3 + \kappa^{-1} A_1 \wedge F_3\eqno(4.19)$$

\noindent
{}From (4.3) we see that the kinetic terms for gravity, dilaton and
antisymmetric
tensors are also correctly described by the generic action $I_{10} (d)$ with $d
= 1, 2, 3$ (i.e $\tilde{d} = 7, 6, 5$).  Both the elementary string ($d = 2$)
and fivebrane $(d = 6)$ solutions of $N = 1$ supergravity described above
continue to provide solutions to Type IIA supergravity, as may be seen by
setting $F_2 = F_4 = 0$.  [This observation is not as obvious as it may seem in
the case of the elementary fivebranes or solitonic strings, however, since it
assumes that one may dualize $F_3$.  Now the Type IIA action follows by
dimensional reduction from the action of $D = 11$ supergravity which contains
$F_4$.  There exists no dual of this action in which $F_4$ is replaced by $F_7$
essentially because $A_3$ appears explicitly in the Chern-Simons term $F_4
\wedge F_4 \wedge A_3$ [16].  Since $F_4$~and~$F_3$ in $D = 10$ originate from
$F_4$ in $D = 11$, this means that we cannot {\it simultaneously} dualize
$F_3$~and~$F_4$ but one may do either {\it separately}.\footnote\dg{We are
grateful to H. Nishino for this observation.}  By partial integration one may
choose to have no explicit $A_3$ dependence in the Chern-Simons term of (4.18)
or no explicity $A_2$ dependence, but not both.]  Furthermore, by setting $F_2
= F_3 = 0$ we find elementary membrane $(d = 3)$ and solitonic fourbrane
$(\tilde{d} = 5)$ solutions, and then by dualizing $F_4$, elementary fourbrane
$(d = 5)$ and solitonic membrane $(\tilde{d} = 3)$ solutions.  Finally, by
setting $F_3 = F_4 = 0$, we find elementary particle $(d = 1)$ and solitonic
sixbrane $(\tilde{d} = 7)$ solutions and then by dualizing $F_2$, elementary
sixbrane $(d = 7)$ and solitonic particle $(\tilde{d} = 1)$ solutions.

Next we consider Type IIB supergravity in $D = 10$ whose bosonic sector
consists of the graviton $g_{MN}$, a complex scalar $\phi$, a complex 2-form
$A_2$ (i.e with $d = 2$ or, by duality $d = 6$) and a real 4-form $A_4$ (i.e
with $d = 4$ which in $D = 10$ is self-dual).  Because of this self-duality of
the 5-form field strength $F_5$, there exists no covariant action principle of
the kind (4.15) and, strictly speaking, our previous analysis ceases to apply.
Nevertheless we can apply the same logic to the equations of motion and we find
that the solution again falls into the generic category (4.8-9).  First of all,
by truncation it is easy to see that the same string $(d = 2)$ and fivebrane
$(d = 6)$ solutions of $N = 1$ supergravity continue to solve the field
equations of Type IIB.  On the other hand, if we set to zero $F_3$ and solve
the self-duality condition $F_5 = - ^{\ast}\!F_5$ then we find the special case
of (4.8) with $d = \tilde{d} = 4$ and hence $\alpha = 0$~and~$\phi = 0$.  This
is the self-dual superthreebrane [6].

All of the above elementary solutions saturate a Bogomoln'yi bound between the
mass per unit $p$-volume ${\cal M}_d$ and the electric charge

$${\cal M}_d = {1\over \sqrt{2}}~\mid e_d \mid\eqno(4.20)$$

\noindent
It follows that the solitonic solutions obey

$${\cal M}_{\tilde{d}} = {1\over \sqrt{2}}~\mid g_{\tilde{d}} \mid\eqno(4.21)$$

\noindent
We shall refer to these equations as the ``mass~$=$~charge'' conditions.
\medskip

\noindent
{\bf 5.  Supersymmetry}
\medskip

Horowitz and Strominger [8] have exhibited a two-parameter family of solutions
of $D = 10$ Type IIA and Type IIB supergravity with event horizons for $d =
1,2,3,4,5,6,7$:  the ``black $p$-branes''.  In some respects, these solutions
resemble the Reissner-Nordstrom black-hole solutions of general relativity
which are known to admit unbroken supersymmetry in the extreme mass~$=$~charge
limit.  Horowitz and Strominger then conjectured that, in this limit, their
black $p$-branes would also be supersymmetric and hence that there exist Type
II super $(d - 1)$ branes for all these values of $d$.  As we shall now
demonstrate, this is indeed the case.

We begin by making the same ansatz as in section 4, namely (4.5-7) but this
time substitute into the supersymmetry transformation rules rather than the
field equations, and demand unbroken supersymmetry.  This reduces the four
unknown functions A, B, C and $\phi$ to one.  We then compare the results
with the known solutions.

For Type IIA supergravity with vanishing fermion background, the gravitino
transformation rule is

$$\eqalign{\delta\psi_M&=D_m \varepsilon + {1\over 64}~e^{3\phi/4}
(\Gamma_M\,^{M_1M_2} - 14 \delta_M\,^{M_1} \Gamma^{M_2}) \Gamma^{11}
\varepsilon F_{M_1M_2}\cr
&\qquad +{1\over 96}~e^{-\phi/2} (\Gamma_M\,^{M_1M_2M_3} - 9 \delta_M\,^{M_1}
\Gamma^{M_2M_3}) \Gamma^{11} \varepsilon F_{M_1M_2M_3}\cr
&\qquad +{i\over 256}~e^{\phi/4} (\Gamma_M\,^{M_1M_2M_3M_4} - {20\over
3}~\delta_M\,^{M_1} \Gamma^{M_2M_3M_4}) \varepsilon
F_{M_1M_2M_3M_4}\cr}\eqno(5.1)$$

\noindent
and the dilatino rule is

$$\eqalign{\delta\lambda&={1\over 4}~\sqrt{2}~D_M \phi \Gamma^M \Gamma^{11}
\varepsilon + {3\over 16}~{1\over \sqrt{2}}~e^{3\phi/4} \Gamma^{M_1M_2}
\varepsilon F_{M_1M_2}\cr
&\qquad \qquad \qquad + {1\over 24}~{i\over \sqrt{2}}~e^{-\phi/2}
\Gamma^{M_1M_2M_3} \varepsilon F_{M_1M_2M_3}\cr
&\qquad \qquad \qquad - {1\over 192}~{i\over \sqrt{2}}~e^{\phi/4}
\Gamma^{M_1M_2M_3M_4} \varepsilon F_{M_1M_2M_3M_4}\cr}\eqno(5.2)$$

\noindent
where $\Gamma^M$ are the $D = 10$ Dirac matrices, where the covariant
derivative is given by

$$D_M = \partial_M + {1\over 4}~\omega_{MAB} \Gamma^{AB}\eqno(5.3)$$

\noindent
with $\omega_{MAB}$ the Lorentz spin connection, where

$$\Gamma^{M_1M_2\ldots M_n} = \Gamma^{[M_1} \Gamma^{M_2} \ldots
\Gamma^{M_n]}\eqno(5.4)$$

\noindent
and where

$$\Gamma^{11} = i \Gamma^0 \Gamma^1 \ldots \Gamma^9\eqno(5.5)$$

\noindent
Similarly the Type IIB rules are

$$\eqalign{\delta \psi_M = D_M \varepsilon&+{i\over 4 \times
480}~\Gamma^{M_1M_2M_3M_4} \Gamma_M \varepsilon F_{M_1M_2M_3M_4}\cr
&+{1\over 96}~(\Gamma_M\,^{M_1M_2M_3} - 9 \delta_M\,^{M_1} \Gamma^{M_2M_3})
\varepsilon^{\ast} F_{M_1M_2M_3}\cr}\eqno(5.6)$$

\noindent
and

$$\delta\lambda = i \Gamma^M \varepsilon^{\ast} P_M - {1\over 24}~i
\Gamma^{M_1M_2M_3} \varepsilon F_{M_1M_2M_3}$$

\noindent
where

$$P_M = \partial_M \phi/(1 - \phi^{\ast} \phi).\eqno(5.7)$$

\noindent
In the Type IIB case, $\varepsilon$ is chiral

$$\Gamma_{11} \varepsilon = \varepsilon\eqno(5.8)$$

The requirement of unbroken supersymmetry is that there exist Killing spinors
$\varepsilon$ for which both $\delta\psi_M$~and~$\delta\lambda$ vanish.
Substituting our ansatze into the transformation rules we find that for every
$1 \leq d \leq 7$ there exist field configurations which break exactly half the
supersymmetries.  This is just what one expects for supersymmetric extended
object solutions [1,7,10,11,12,13] and is intimately related to the
$\kappa$-symmetry discussed in section 2 and the Bogomoln'yi bounds of section
4.  The corresponding values of A, B and $\phi$ in terms of C are given in
Table 2.  The important observation, from (4.8), is that the values required by
supersymmetry also solve the field equations.  Thus in addition to the $D = 10$
super $(d - 1)$ branes already known to exist for $d = 2$ (Heterotic, Type IIA
and Type IIB), $d = 4$ (Type IIB only) and $d = 6$ (Heterotic, Type IIA and
Type IIB), we have established the existence of a Type IIA superparticle $(d =
1)$, a Type IIA supermembrane $(d = 3)$, a Type IIA superfourbrane $(d = 5)$
and a Type IIA supersixbrane $(d = 7)$.

One may now repeat the $D = 10$ analysis of sections 4 and 5 for $N = 2$
supergravities in $D < 10$.  Here we simply state the results.  Details will be
discussed elsewhere [17].  We find supersymmetric solitons for all $1 \leq
\tilde{d} \leq 7$ where $\tilde{d} = D - 2 - d$, as shown in Fig. 3.  At first
sight, this seems to contradict Fig. 2 since solutions appear where no
supermultiplet is allowed.  The resolution is simply that only the cases $d =
1, 3, 4, 5, 6$~and~$7$ in $D = 10$ are fundamental.  All the others are
obtained by simply dimensional reduction of these or the $D = 11$
supermembrane, and are thus described by the same gauge-fixed action.  In
summary the new brane-scan including (fundamental) Type II super $p$-branes is
given in Fig. 4.
\medskip

\noindent
{\bf 6.  Conclusions}
\medskip

We have classified all supersymmetric extended objects that correspond to
solitons of a Poincar\'e supersymmetric field theory in the usual spacetime
signature which break half the spacetime supersymmetries, as shown in Fig. 4.
We cannot at the present time rigorously rule out the existence of other super
$p$-branes, denoted by the points in Fig. 2 not appearing in Fig. 4, which do
not correspond to solitons.  However, we regard their existence as unlikely.
(Nor can we rule out the possibility of other super $p$-branes described by
non-Poincar\'e supersymmetries in other signatures as discussed in [18] e.g a
(2,2) worldvolume in a (10,2) spacetime).  Further progress would require that
we construct the spacetime Green-Schwarz supersymmetric and $\kappa$-symmetric
actions for these new Type II $p$-branes and, to date, this has not been done.
All we know is that, in a physical gauge, the worldvolume theory corresponding
to the zero modes of the soliton is described by vector or antisymmetric tensor
supermultiplet as in Table 3.
\bigskip

\halign{\indent #&\hfil # \hfil &\quad \hfil # \hfil &\quad \hfil # \hfil &
\quad \hfil# \hfil &\quad \hfil # \hfil&\quad #\hfil \cr
&$d = 7$ & Type IIA & $(A_{\mu}, \lambda, 3\phi)$ &  & $n = 1$&\cr
&$d = 6$ & Type IIA & $(B_{\mu\nu}^-, \lambda^I, \phi^{[IJ]})$ & $I = 1,
\ldots, 4$ & $(n_+, n_-) = (2,0)$&\cr
&   & Type IIB & $(B_{\mu}, \chi^I, A^I\,_J, \xi)$ & $I = 1, 2$ & $(n_+, n_-)
= (1,1)$&\cr
&$d = 5$ & Type IIA & $(A_{\mu}, \lambda^I, \phi^{[IJ]\mid})$ & $I = 1, \ldots,
4$ & $n = 2$&\cr
&$d = 4$ & Type IIB & $(B_{\mu}, \chi^I, \phi^{[IJ]})$ & $I = 1, \ldots, 4$ &
$n = 4$&\cr
&$d = 3$ & Type IIA & $(\chi^I, \phi^I)$ & $I = 1, \ldots, 8$ & $n = 8$&\cr
&$d = 2$ & Type IIA & $(\lambda_L\,^I, \phi_L\,^I)$ & $I = 1, \ldots, 16$ &
$(n_+, n_-) = (16,0)$&\cr
&      & Type IIB & $(\chi^I_L, \phi_L\,^I), (\chi^I_R, \phi^I_R)$ & $I = 1,
\ldots, 8$ & $(n_+, n_-) = (8,8)$&\cr}

\itemitem{Table 3:} Gauge-fixed theories on the worldvolume, corresponding to
the zero modes of the soliton, are described by the above supermultiplets.
\medskip

The case of the Type IIA membrane in $D = 10$ is particularly interesting.  It
emerges as an elementary solution of the usual formulation of Type IIA
supergravity with a 4-form field strength or else as a soliton solution of the
dual formulation with a 6-form field strength.  Indeed, this solution is dual
to the $D = 10$ superfourbrane solution.  However, its zero-modes are on-shell
equivalent to those of the $D = 11$ supermembrane and so does not occupy a
separate slot on the brane-scan of Fig. 3.  The reason is because in $d = 3$
the worldvolume vector has only 1 degree of freedom and is dual (in the
three dimensional sense) to a scalar.  Indeed, this provides the exception to
the rule that we do not have an explicit expression for the Type II $p$-brane
Green-Schwarz actions for $p > 1$.  Its action is obtained by making a $10 + 1$
split of the Green-Schwarz action for the $D = 11$ supermembrane coordinates
$\hat{X}^{\hat{M}} = (X^M, X^{10})$ where $\hat{M} = 0,1,\ldots 10$~and~$M =
0,1,\ldots 9$, and then dualizing the $X^{10}$.  Perhaps the most bizarre
aspect of all this is that an object living in eleven dimensions should emerge
as a soliton of a ten dimensional theory!

Of course, one might ask why the $D = 7$ membrane occupies a separate slot
since it too can be viewed as a spacetime dimensional reduction of the $D = 11$
membrane.  The answer is that when we reach $D = 7$ the multiplet becomes
reducible and we can thus perform a consistent truncation to a smaller theory
with half the supersymmetries.  Similar remarks apply to the membranes in $D =
5$~and 4, and indeed to all the circles appearing on the brane-scan.

In our classification, we have also omitted supersymmetric solitons which break
{\it more than half} the supersymmetries since these solutions presumably admit
no $\kappa$-symmetric Green-Schwarz action (at least, not of the kind
presently known).  Examples of this are provided by the $D = 10$ octonionic
string of Harvey and Strominger [19], (which breaks 7/8), and the $D = 11$
extreme black fourbrane and extreme black sixbrane of G\"uven [20] (which break
3/4 and 7/8, respectively).

Finally, we ask what are the implications of our results for the idea of
``duality'', in the sense that one theory is simply providing a dual
description of the same physics of another theory [21] with the weak-coupling
regime of one being the strong-coupling of the other [11]?  At the classical
level discussed in this paper, we see that supersymmetry has narrowed down the
possibilities to just four, all in $D = 10$, namely particle/sixbrane duality
(Type IIA only), string/fivebrane duality (Heterotic, Type IIA or Type IIB),
membrane/fourbrane duality (Type IIA only) and threebrane self-duality (Type
IIB only).  The implications for quantum duality will be discussed elsewhere.
\vfill\eject

\noindent
{\bf Acknowledgements}
\medskip

We have enjoyed useful conversations with E. Bergshoeff, R. Khuri, E. Sezgin,
K. Stelle and P.  Townsend.
\vfill\eject

\centerline{\bf REFERENCES}
\medskip

\item{1.} J. Hughes, J. Liu and J. Polchinski, Phys. Lett. {\bf B180} (1986)
370.

\item{2.} E. Bergshoeff, E. Sezgin and P. K. Townsend, Phys. Lett. {\bf B189}
(1987) 75.

\item{3.} M. J. Duff, P. Howe, T. Inami, and K. S. Stelle, Phys. Lett. {\bf
B191} (1987) 70.

\item{4.} A. Achucarro, J. M. Evans, P. K. Townsend and D. L. Wiltshire, Phys.
Lett. {\bf B198} (1987) 441.

\item{5.} C. G. Callan, J. A. Harvey and A. Strominger, Nucl. Phys. {\bf B359}
(1991) 611; Nucl. Phy. {\bf B367} (1991) 60.

\item{6.} M. J. Duff and J. X. Lu, Phys. Lett. {\bf B273} (1991) 409.

\item{7.} P. K. Townsend, Phys. Lett. {\bf B202} (1988) 53.

\item{8.} G. T. Horowitz and A. Strominger, Nucl. Phys. {\bf B360} (1991) 197.

\item{9.} J. Strathdee, Int. J. Mod. Phys. {\bf A2} (1987) 273.

\item{10.} A. Dabholkar, G. Gibbons, J. Harvey and F. Ruiz Ruiz, Nucl. Phys.
{\bf B340} (1990) 33.

\item{11.} A. Strominger, Nucl. Phys. {\bf B343} (1990) 167.

\item{12.} M. J. Duff and K. S. Stelle, Phys. Lett. {\bf B253} (1991) 113.

\item{13.} M. J. Duff and J. X. Lu, Nucl. Phys. {\bf B354} (1991) 141.

\item{14.} M. J. Duff and J. X. Lu, Phys. Rev. Lett. {\bf 66} (1991) 1402.

\item{15.} M. J. Duff and J. X. Lu, Class. Quantum Grav. {\bf 9} (1992) 1.

\item{16.} H. Nicolai, P. K. Townsend and P. van Nieuwenhuizen, Lett. Nuovo
Cimento {\bf 30} (1981) 315.

\item{17.} M. J. Duff and J. X. Lu, ``Black and super $p$-branes in diverse
dimensions'', CTP-TAMU-54/92.

\item{18.} M. Blencowe and M. J. Duff, Nucl. Phys. {\bf B310} (1988) 387.

\item{19.} J. Harvey and A. Strominger, Phys. Rev. Lett. {\bf 66} (1991) 549.

\item{20.} R. G\"uven, ``Black $p$-brane solutions of $D = 11$ Supergravity
Theory,'' Bo\u gazici University preprint (1991).

\item{21.} M. J. Duff, Class. Quantum Grav. {\bf 5} (1988) 189.
\vfill\eject\bye